\documentclass[12pt]{JHEP3}

\title{Entropy Function for Non-Extremal Black Holes in String Theory}
\author{Rong-Gen Cai \\
        Institute of Theoretical Physics\\
    Chinese Academy of Sciences\\
    P.O.Box 2735, Beijing 100080, China\\
    \email{cairg@itp.ac.cn}}
\author{Da-Wei Pang\\
        Institute of Theoretical Physics\\
    Chinese Academy of Sciences\\
    P.O.Box 2735, Beijing 100080, China\\
    {\rm and}\\
    Graduate University of the Chinese Academy of Sciences\\
    YuQuan Road 19A, Beijing 100049, China\\
    \email{pangdw@itp.ac.cn}}

\abstract{We generalize the entropy function formalism to
five-dimensional and four-dimensional non-extremal black holes in
string theory. In the near horizon limit, these black holes have BTZ
metric as part of the spacetime geometry. It is shown that the
entropy function formalism also works very well for these
non-extremal black holes and it can reproduce the Bekenstein-Hawking
entropy of these black holes in ten dimensions and lower
dimensions.}

\begin{document}

\section{Introduction}
\label{Introduction}
The black hole attractor mechanism has been an interesting subject
over the past years, which states that in a black hole background
the moduli fields vary radially and get ``attracted'' to certain
specific values at the horizon which depend only on the quantized
charges of the black hole under consideration. As a result, the
macroscopic entropy of the black hole is given only in terms of the
charges and is independent of the asymptotic values of the moduli.
It was first discovered in the context of N=2 extremal black
holes~\cite{mc,classic} and was generalized to theories with higher
derivative corrections in~\cite{cardoso}. The attractor mechanism
for non-supersymmetric black holes was initiated in~\cite{fgk} and
discussed more extensively in following
papers~\cite{non},~\cite{kal} and~\cite{dst}.

Recently, Sen proposed an efficient way to calculate the entropy of
an extremal black hole, which is named as ``entropy function''
formalism~\cite{sen}. The main steps can be summarized as follows:

i) Define a $d$-dimensional extremal black hole to be such an object
that the near horizon geometry is given by $AdS_{2}\times S^{d-2}$.
Choose a coordinate system in which the $AdS_{2}$ part of the metric
is proportional to $-r^2dt^2+dr^2/r^2$. The black hole background is
supported by electric and magnetic fields, as well as various moduli
scalar fields.

ii) Consider a general $AdS_{2}\times S^{d-2}$ background
characterized by the sizes of $AdS_{2}$ and $S^{d-2}$, the electric
and magnetic fields and various scalar fields. Define an entropy
function by carrying an integral of the Lagrangian density over
$S^{d-2}$ and then taking the Legendre transform of the integral
with respect to the parameters $e_{i}$ denoting the electric fields.
The result is a function of the moduli values $u_{s}$, the sizes
$v_{1}$ and $v_{2}$ of $AdS_{2}$ and $S^{d-2}$, the electric charges
$q_{i}$ conjugate to $e_{i}$, and the magnetic charges $p_{a}$.

iii) For given electric and magnetic charges $\{q_{i}\}$ and
$\{p_{a}\}$, the values $u_{s}$ of the scalar fields as well as the
sizes $v_{1}$ and $v_{2}$ are determined by extremizing the entropy
function with respect to the variables $u_{s}$, $v_{1}$ and $v_{2}$.
Finally the entropy is given by the value of the entropy function at
the horizon.

This is a very simple and powerful method to calculate entropy of
such kind of black holes.  In particular, one can easily obtain the
corrections to the entropy due to higher order corrections in the
effective Lagrangian. Several related works are given
in~\cite{relate}.

As is well known that in string theory, some kinds of black holes
can be constructed by putting D-branes together and the
Bekenstein-Hawking entropy can be understood by counting the
degeneracies of the microstates of such configurations. Such
extremal black holes have $AdS_{3}$ as part of the near horizon
geometry in ten dimensions instead of $AdS_{2}$. After dimensional
reduction down to lower dimensions, the near horizon geometry turns
out to be $AdS_{2}$ times a sphere. The entropy function for $D1D5P$
extremal black hole in Type IIB string theory and $D2D6NS5P$
extremal black hole in Type IIA were calculated in~\cite{ag}
and~\cite{cp}, where it was shown that the entropy function
formalism could give the correct entropy both in ten dimensions and
lower dimensions.

However, for some non-extremal black holes constructed by D-branes,
part of the near horizon geometry turns out to be BTZ black hole.
Since BTZ black hole is locally equivalent to $AdS_{3}$, one expects
that the entropy function formalism is also applicable for those
black holes. In this paper we show it is indeed the case: after
taking the near horizon limit, the entropy function for 5$d$
non-extremal black hole can give the Bekenstein-Hawking entropy
precisely, while the entropy function for 4$d$ non-extremal black
hole gives the Bekenstein-Hawking entropy up to a  factor, which can
be understood via a rescaling transformation relation of entropy
function.

The rest of the paper is organized as follows: In
section~\ref{Non-ext} we review the basic properties of the
non-extremal black holes. We give a general proof of the entropy
function formula for black hole with BTZ as part of its near horizon
geometry in section~\ref{General}. We calculate the entropy function
for non-extremal black holes in ten dimensions and lower dimensions
in section~\ref{ten-dim} and section~\ref{lower} respectively.
Finally in the last section~\ref{Summary} we summarize our results
and discuss related topics.

\section{Non-extremal Black Holes in String Theory}
\label{Non-ext}
 It is known that non-extremal black holes can be constructed
 by putting D-branes together, so that the entropy can also be
 obtained by counting the degrees of freedom of the D-brane system
 after taking near-extremal limit~\cite{near}. The microscopic entropy
 of such kinds of black holes can also be understood via
 U-duality~\cite{udual}. In this section we make a brief review for the
 salient properties of non-extremal black holes which are necessary in
 the following calculations. For reviews, see~\cite{rev}.
\subsection{The 5d non-extremal black hole}
The five dimensional non-extremal black hole can be constructed by
using $D1$-branes, $D5$-branes and momentum $P$, which is a solution
of Type IIB supergravity. The effective action is
\begin{equation}
S=\frac{1}{16\pi G^{10}_{N}}\int d^{10}x\sqrt{-{\rm
det}~g}\{e^{-2\phi}[R+4(\nabla\phi)^2]-\frac{1}{2}\sum\limits_{n}\frac{1}{n!}F^{2}_{n}\},
\label{iib}
\end{equation}
where $F_{n}$ denote the field strengths carried by the $D$-branes.

The non-extremal black hole metric in $d=10$ is given as follows in
string frame:
\begin{eqnarray}
ds^{2}_{10}&=&f_{1}(r)^{-\frac{1}{2}}f_{5}(r)^{-\frac{1}{2}}[-dt^{2}+dz^{2}+K(r)(\cosh
              \alpha_{m}dt-\sinh\alpha_{m}dz)^2] \nonumber \\
           & &+f_{1}(r)^{\frac{1}{2}}f_{5}(r)^{-\frac{1}{2}}dx^{2}_{\|}+f_{1}(r)^{\frac{1}{2}}f_{5}(r)^{\frac{1}{2}}
           [\frac{dr^{2}}{1-K(r)}+r^2d\Omega^{2}_{3}],\nonumber \\
e^{-2\phi} &=&\frac{f_{5}}{f_{1}},
\end{eqnarray}
where
\begin{equation}
K(r)=\frac{r_{H}^{2}}{r^{2}},~~~f_{1}(r)\equiv1+\frac{r_{1}^{2}}{r^{2}}=1+\frac{r_{H}^{2}\sinh^{2}\alpha_{1}}{r^{2}},
~~~f_{5}(r)\equiv1+\frac{r_{5}^{2}}{r^{2}}=1+\frac{r_{H}^{2}\sinh^{2}\alpha_{5}}{r^{2}},
\end{equation}
and $\alpha$'s are the boost parameters. The $D1$-branes can be
viewed as the electric source carrying 3-form electric field
strength while the $D5$-branes can be regarded as the magnetic
source carrying dual 3-form magnetic field strength. The conserved
charges are given by
\begin{equation}
Q_{1}=\frac{Vr_{H}^{2}}{g_{s}}\frac{\sinh(2\alpha_{1})}{2},~~~Q_{5}=\frac{r_{H}^{2}}{g_{s}}\frac{\sinh(2\alpha_{5})}{2},
~~~N=\frac{R_{z}^{2}Vr_{H}^{2}}{g_{s}^{2}}\frac{\sinh(2\alpha_{m})}{2},
\end{equation}
where the fundamental string length $l_{s}$ has been taken to be 1.
Here $V=R_{5}R_{6}R_{7}R_{8}$ where $R_{i},~i=5,6,7,8$ denote the
radii of the four coordinates in $x_{\|}$ and $R_{z}$ is the radius
of the compact dimension $z$, along which there is a momentum $P$.
The Bekenstein-Hawking entropy is
\begin{equation}
S_{BH}=\frac{2\pi
R_{z}Vr_{H}^{3}}{g_{s}^{2}}(\cosh\alpha_{1}\cosh\alpha_{5}\cosh\alpha_{m}).
\end{equation}

To obtain the near horizon geometry, we take the limit
\begin{equation}
r^{2}\ll r_{1,5}^{2}\equiv r_{H}^{2}\sinh^{2}\alpha_{1,5},
\end{equation}
but we do not demand a similar condition on $r_{m}\equiv
r_{H}\sinh\alpha_{m}$. This limit means that
$\alpha_{1},~\alpha_{5}$ tend to be very large in the near horizon
region, so that $\sinh\alpha_{1,5}\approx\cosh\alpha_{1,5}$. Note
that the near horizon limit is just the decoupling limit in the
AdS/CFT correspondence. In addition,  when $r_{H}\rightarrow0$ and
$\alpha_{1},~\alpha_{5}\rightarrow\infty$  while keeping the charges
fixed, the non-extremal black hole turns out to be extremal. After
taking such a near horizon limit, the metric becomes
\begin{eqnarray}
ds^{2}&=&-\frac{(\rho^{2}-\rho_{+}^{2})(\rho^{2}-\rho_{-}^{2})}{\lambda^{2}\rho^{2}}dt^{2}+\frac{\lambda^{2}\rho^{2}}
{(\rho^{2}-\rho_{+}^{2})(\rho^{2}-\rho_{-}^{2})}d\rho^{2}\nonumber
\\
      & &+\rho^{2}(dy-\frac{\rho_{+}\rho_{-}}{\lambda\rho^{2}}dt)^2+\lambda^{2}d\Omega_{3}^{2}+\frac{r_{1}}{r_{5}}dx_{\|}^{2},
\end{eqnarray}
which is of the form $BTZ\times S^{3}\times T^{4}$. Note that here
we have made the coordinate transformation
\begin{equation}
\label{res1} \rho^{2}\equiv r^{2}+\rho_{-}^{2},~~~y\equiv
\frac{z}{\lambda}
\end{equation}
and have introduced the parameters
\begin{equation}
\rho_{+}\equiv r_{H}\cosh\alpha_{m},~~~\rho_{-}\equiv
r_{H}\sinh\alpha_{m},~~~\lambda^{2}\equiv r_{1}r_{5}.
\end{equation}
\subsection{The 4d non-extremal black hole}
The four dimensional non-extremal black hole can be taken as a
non-extremal intersection of $D2$-branes in $(z, x_{2})$,
$D6$-branes in $(z, x_{2}, x_{3}, x_{4}, x_{5}, x_{6})$,
$NS5$-branes in $(z,x_{3}, x_{4}, x_{5}, x_{6})$ and momentum $P$
along $z$, which is a solution of type IIA supergravity with the
effective action
\begin{eqnarray}
S=\frac{1}{16\pi G^{10}_{N}}\int d^{10}x\sqrt{-{\rm
det}~g}&&\{e^{-2\phi}[R+4(\nabla\phi)^{2}-\frac{1}{3}H^{2}]-G^{2}-\frac{1}{12}F^{\prime
2} \nonumber \\
       &&-\frac{1}{288}\epsilon^{\mu_{1}\cdots\mu_{10}}F_{\mu_{1}\mu_{2}\mu_{3}\mu_{4}}
 F_{\mu_{5}\mu_{6}\mu_{7}\mu_{8}}B_{\mu_{9}\mu_{10}}\},
\label{iia}
\end{eqnarray}
where the 2-form $G=dA$, 3-form $H=dB$, 4-form $F=dC$ are the field
strengths carried by D6-branes, NS5-branes, D2-branes respectively
and $F^{\prime}=F+2A\wedge H$.

The four dimensional non-extremal black hole metric, written in ten
dimensional string frame, is given as follows:
\begin{eqnarray}
ds^{2}_{10}&=&(f_{2}f_{6})^{-\frac{1}{2}}[-K^{-1}fdt^{2}+K(dz+(K^{\prime-1}-1)dt)^{2}]
\nonumber \\
           & &+f_{5}(f_{2}f_{6})^{-\frac{1}{2}}dx^{2}_{2}+f_{2}^{\frac{1}{2}}f_{6}^{-\frac{1}{2}}(dx_{3}^{2}+dx_{4}^{2}+dx_{5}^{2}+dx_{6}^{2})
          \nonumber \\
           &
           &+f_{5}(f_{2}f_{6})^{\frac{1}{2}}(f^{-1}dr^{2}+r^{2}d\Omega_{2}^{2}),
           \nonumber \\
 e^{-2\phi}&=&f_{5}^{-1}f_{6}^{\frac{3}{2}}f_{2}^{-\frac{1}{2}},
\end{eqnarray}
where
\begin{eqnarray}
f_{2}&\equiv&1+\frac{r_{2}}{r}=1+\frac{r_{H}\sinh^{2}\alpha_{2}}{r},~~~f_{5}\equiv1+\frac{r_{5}}{r}=1+\frac{r_{H}\sinh^{2}\alpha_{5}}{r}, \nonumber\\
f_{6}&\equiv&1+\frac{r_{6}}{r}=1+\frac{r_{H}\sinh^{2}\alpha_{6}}{r},~~~K\equiv1+\frac{r_{K}}{r}=1+\frac{r_{H}
\sinh^{2}\alpha_{K}}{r},\nonumber
\\
K^{\prime-1}&=&1-\frac{q_{K}}{r}K^{-1}=1-\frac{r_{H}\sinh\alpha_{K}\cosh\alpha_{K}}{r}K^{-1},~~~f=1-\frac{r_{H}}{r}.
\end{eqnarray}
Here the $D2$-branes can be taken as the electric source carrying
4-form electric field strength while the $D6$-branes and the
$NS5$-branes can be taken as the magnetic source carrying dual
2-form and 3-form magnetic field strengths respectively. Then we can
obtain the conserved charges
\begin{eqnarray}
Q_{2}&=&\frac{r_{H}V}{g_{s}}\sinh2\alpha_{2},~~~Q_{5}=r_{H}R_{2}\sinh2\alpha_{5},\nonumber
\\
Q_{6}&=&\frac{r_{H}}{g_{s}}\sinh2\alpha_{6},~~~N=\frac{r_{H}VR_{z}^{2}R_{2}}{g_{s}^{2}}\sinh2\alpha_{K},
\end{eqnarray}
where the fundamental string length $l_{s}$ has been set to be 1.
Here $V=R_{3}R_{4}R_{5}R_{6}$ where $R_{i},i=2,3,4,5,6$ denote the
radii of the coordinates $x_{i}$ and $R_{z}$ is the radius of the
compact dimension $z$. The Bekenstein-Hawking entropy is
\begin{equation}
S_{BH}=\frac{8\pi
r_{H}^{2}VR_{2}R_{z}}{g_{s}^{2}}\cosh\alpha_{2}\cosh\alpha_{5}\cosh\alpha_{6}\cosh\alpha_{K}.
\end{equation}

The near horizon geometry can be obtained in a similar way. First we
take the near horizon limit, i.e. we require that
\begin{equation}
r\ll r_{2,5,6}\equiv r_{H}\sinh^{2}\alpha_{2,5,6},
\end{equation}
but we do not demand a similar condition on $r_{K}\equiv
r_{H}\sinh^{2}\alpha_{K}$. This limit means that
$\alpha_{2},~\alpha_{5},~\alpha_{6}$ tend to be very large when the
near horizon region is approached, so that
$\sinh\alpha_{2,5,6}\approx\cosh\alpha_{2,5,6}$. Note that when
$r_{H}\rightarrow0$ and
$\alpha_{2},~\alpha_{5},~\alpha_{6}\rightarrow\infty$ while keeping
the charges fixed, the non-extremal black hole turns out to be
extremal. After taking near horizon limit, the  metric becomes
\begin{eqnarray}
ds^{2}&=&-\frac{(\rho^{2}-\rho_{+}^{2})(\rho^{2}-\rho_{-}^{2})}{\lambda^{2}\rho^{2}}d\tau^{2}+\frac{\lambda^{2}\rho^{2}}
{(\rho^{2}-\rho_{+}^{2})(\rho^{2}-\rho_{-}^{2})}d\rho^{2} \nonumber
\\
      & &+\rho^{2}(dy-\frac{\rho_{+}\rho_{-}}{\lambda\rho^{2}}d\tau)^2+\frac{\lambda^{2}}{4}d\Omega_{2}^{2}
      \nonumber \\
      & &+\frac{r_{5}}{(r_{2}r_{6})^{\frac{1}{2}}}dx_{2}^{2}+(\frac{r_{2}}{r_{6}})^{\frac{1}{2}}(dx_{3}^{2}+dx_{4}^{2}+dx_{5}^{2}+dx_{6}^{2}),
\end{eqnarray}
which is of the form $BTZ\times S^{2}\times S^{1}\times T^{4}$. Note
that here we have made the coordinate transformation
\begin{equation}
\label{res2}
\tau=2\sqrt{r_{5}}t,~~~y=\frac{z}{(r_{2}r_{6})^{\frac{1}{4}}},~~~\rho^{2}=r+\rho_{-}^{2}
\end{equation}
and have introduced the parameters
\begin{equation}
\rho_{+}^{2}\equiv r_{H}\cosh^{2}\alpha_{K},~~~\rho_{-}^{2}\equiv
r_{H}\sinh^{2}\alpha_{K},~~~\lambda^{2}\equiv4r_{5}\sqrt{r_{2}r_{6}}.
\end{equation}
\section{The Entropy Function Formalism: General Proof}
\label{General}
In this section we will give a detailed derivation of the entropy
function for non-extremal black holes with $BTZ$ as part of the near
horizon geometry, following~\cite{ag} and~\cite{cp}. Note that the
entropy function formalism originates from Wald's entropy
formula~\cite{wald}, which requires that the black hole under
consideration should have a bifurcate horizon. Then in the entropy
function formalism, the entropy of an extremal black hole should be
taken as the extremal limit of a non-extremal black hole. So it is
natural to expect that the entropy function formalism is also
applicable to non-extremal black holes.

A generalized form of Wald formula was proposed in~\cite{mv}, which
states that
\begin{equation}
\mathcal{S}_{BH}=4\pi\int_{H}dx_{H}\sqrt{{\rm
det}g_{H}}\frac{\partial\mathcal{L}}{\partial
R_{\mu\nu\lambda\rho}}g^{\bot}_{\mu\lambda}g^{\bot}_{\nu\rho},
\label{wald}
\end{equation}
where $\mathcal{L}$ is the Lagrangian density, ${\rm det}g_{H}$ is
the determinant of the horizon metric and $g^{\bot}_{\mu\nu}$
denotes the orthogonal metric obtained by projecting onto subspace
orthogonal to the horizon. For a metric of the general form
\begin{equation}
ds^{2}=g_{tt}dt^{2}+g_{yy}dy^{2}+2g_{ty}dtdy+g_{rr}dr^{2}+d\vec{x}^{2},
\end{equation}
the orthogonal metric is defined as
\begin{equation}
g^{\bot}_{\mu\nu}=(N_{t})_{\mu}(N_{t})_{\nu}+(N_{r})_{\mu}(N_{r})_{\nu},
\end{equation}
where $N_{t}$ and $N_{r}$ are unit normal vectors to the horizon
\begin{equation}
N_{t}=\sqrt{\frac{g^{yy}}{g^{tt}g^{yy}-(g^{ty})^{2}}}(1,0,-\frac{g^{ty}}{g^{yy}},0),~~~
N_{r}=(0,\frac{1}{\sqrt{g^{rr}}},0,0).
\end{equation}

Consider the general near horizon metric which has a BTZ part
\begin{eqnarray}
ds^{2}&=&v_{1}[-\frac{(\rho^{2}-\rho_{+}^{2})(\rho^{2}-\rho_{-}^{2})}{\lambda^{2}\rho^{2}}d\tau^{2}
+\frac{\lambda^{2}\rho^{2}}{(\rho^{2}-\rho_{+}^{2})(\rho^{2}-\rho_{-}^{2})}d\rho^{2}\\\nonumber
      &
      &+\rho^{2}(dy-\frac{\rho_{+}\rho_{-}}{\lambda\rho^{2}}d\tau)^2]+v_{2}d\vec{x}^{2}.
\end{eqnarray}
The relevant orthogonal metric and Riemann tensor components are
given below
\begin{eqnarray}
g^{\bot}_{\tau\tau}&=&\frac{(\rho_{+}^{2}+\rho_{-}^{2}-\rho^{2})v_{1}}{\lambda^{2}},~~~
g^{\bot}_{\tau y}=-\frac{\rho_{+}\rho_{-}v_{1}}{\lambda},\nonumber
\\
      g^{\bot}_{yy}&=&\frac{\rho_{+}^{2}\rho_{-}^{2}v_{1}}{(\rho_{+}^{2}+\rho_{-}^{2}-\rho^{2})},~~~
g^{\bot}_{\rho\rho}=\frac{\lambda^{2}\rho^{2}v_{1}}{(\rho^{2}-\rho_{+}^{2})(\rho^{2}-\rho_{-}^{2})},
\end{eqnarray}
\begin{eqnarray}
R_{\tau\rho\tau\rho}&=&\frac{\rho^{2}(\rho^{2}-\rho_{+}^{2}-\rho_{-}^{2})v_{1}}
{\lambda^{2}(\rho^{2}-\rho_{+}^{2})(\rho^{2}-\rho_{-}^{2})},\nonumber
\\
     R_{y\rho
     y\rho}&=&-\frac{\rho^{4}v_{1}}{(\rho^{2}-\rho_{+}^{2})(\rho^{2}-\rho_{-}^{2})},\nonumber
     \\
  R_{\tau\rho y\rho}&=&\frac{\rho^{2}\rho_{+}\rho_{-}v_{1}}{\lambda(\rho^{2}-\rho_{+}^{2})(\rho^{2}-\rho_{-}^{2})}.
\end{eqnarray}
Then the Wald formula~(\ref{wald}) can be rewritten as
\begin{equation}
\mathcal{S}_{BH}=\sum\limits_{i=1}^{4}\mathcal{S}_{i}, \label{s}
\end{equation}
where
\begin{eqnarray}
\label{si} \mathcal{S}_{1}&=&8\pi\int_{H}dx_{H}\sqrt{{\rm
det}g_{H}}\frac{\partial\mathcal{L}}{\partial
R_{\tau\rho\tau\rho}}g^{\bot}_{\tau\tau}g^{\bot}_{\rho\rho}
\nonumber \\
               &=&-8\pi\lambda^{2}v_{1}\int_{H}\sqrt{{\rm
det}g_{H}}\frac{\partial\mathcal{L}}{\partial
R_{\tau\rho\tau\rho}}R_{\tau\rho\tau\rho}, \nonumber \\
\mathcal{S}_{2}&=&8\pi\int_{H}dx_{H}\sqrt{{\rm
det}g_{H}}\frac{\partial\mathcal{L}}{\partial R_{y\rho
y\rho}}g^{\bot}_{yy}g^{\bot}_{\rho\rho} \nonumber \\
               &=&8\pi\frac{\lambda^{2}\rho_{+}^{2}\rho_{-}^{2}v_{1}}{\rho^{2}(\rho^{2}-\rho_{+}^{2}-\rho_{-}^{2})}\int_{H}\sqrt{{\rm
det}g_{H}}\frac{\partial\mathcal{L}}{\partial R_{y\rho
y\rho}}R_{y\rho y\rho}, \nonumber \\
\mathcal{S}_{3}&=&16\pi\int_{H}dx_{H}\sqrt{{\rm
det}g_{H}}\frac{\partial\mathcal{L}}{\partial R_{\tau\rho
y\rho}}g^{\bot}_{\tau y}g^{\bot}_{\rho\rho} \nonumber \\
               &=&-16\pi\lambda^{2}v_{1}\int_{H}\sqrt{{\rm
det}g_{H}}\frac{\partial\mathcal{L}}{\partial R_{\tau\rho
y\rho}}R_{\tau\rho y\rho}, \nonumber \\
\mathcal{S}_{4}&=&8\pi\int_{H}dx_{H}\sqrt{{\rm
det}g_{H}}\frac{\partial\mathcal{L}}{\partial R_{\tau y\tau
y}}(g^{\bot}_{\tau\tau}g^{\bot}_{yy}-(g^{\bot}_{\tau y})^{2})
\nonumber \\
               &=&0.
\end{eqnarray}

Next we define a function $f$ as integral of the Lagrangian density
over the horizon,
\begin{equation}
\label{3eq10} f\equiv\int_{H}dx_{H}\sqrt{-{\rm det}g}\mathcal{L}
\end{equation}
and rescale the Riemann tensor components as proposed in~\cite{sen},
\begin{eqnarray}
R_{\tau\rho\tau\rho}&\rightarrow&\lambda_{1}R_{\tau\rho\tau\rho},~~~
R_{\rho y\rho y}\rightarrow\lambda_{2}R_{\rho y\rho y}, \nonumber
\\
R_{\tau\rho y\rho}&\rightarrow&\lambda_{3}R_{\tau\rho y\rho},~~~
R_{\tau y\tau y}\rightarrow\lambda_{4}R_{\tau y\tau y}.
\end{eqnarray}
It can be seen that the rescaled Lagrangian $\mathcal{L}_{\lambda}$
behaves as
\begin{equation}
\frac{\partial\mathcal{L}_{\lambda}}{\partial\lambda_{i}}=R^{(i)}_{\mu\nu\lambda\rho}
\frac{\partial\mathcal{L}_{\lambda}}{\partial
R^{(i)}_{\mu\nu\lambda\rho}},~~~i=1,2,3,4
\end{equation}
Then the rescaled function $f_{\lambda}$ satisfies the following
relation
\begin{equation}
\label{3eq13}
 \left. \frac{\partial
f_{\lambda}}{\partial\lambda_{i}} \right
|_{\lambda_{i}=1}=v_{1}\int_{H}dx_{H}\sqrt{{\rm
det}g_{H}}R^{(i)}_{\mu\nu\lambda\rho}
\frac{\partial\mathcal{L}_{\lambda}}{\partial
R^{(i)}_{\mu\nu\lambda\rho}}
\end{equation}
with no summation on the right hand side for $i$. Substituting these
relations into~(\ref{s}) and~(\ref{si}), we obtain the following
expression for $\mathcal{S}_{BH}$
\begin{equation}
\label{3eq14} \mathcal{S}_{BH}=-2\pi\lambda^{2}
\left.(\frac{\partial
f_{\lambda_{1}}}{\partial\lambda_{1}}+\frac{\rho_{+}^{2}\rho_{-}^{2}}{\rho^{2}(\rho_{+}^{2}+\rho_{-}^{2}-\rho^{2})}
\frac{\partial f_{\lambda_{2}}}{\partial\lambda_{2}}+\frac{\partial
f_{\lambda_{3}}}{\partial\lambda_{3}}) \right
|_{\lambda_{1}=\lambda_{2}=\lambda_{3}=1}.
\end{equation}
Furthermore, since the general Lagrangian should be diffeomorphism
invariant, the components of the Riemann tensor entering the
Lagrangian must be accompanied by the corresponding components of
the inverse metric. Thus we have the following relations
\begin{eqnarray}
\lambda_{1}R_{\tau\rho\tau\rho}g^{\tau\tau}g^{\rho\rho}&\sim&\lambda_{1}v_{1}^{-1},~~~
\lambda_{2}R_{\rho y\rho
y}g^{yy}g^{\rho\rho}\sim\lambda_{2}v_{2}^{-1}, \nonumber \\
\lambda_{3}R_{\tau\rho y\rho}g^{\tau y}g^{\rho\rho}
&\sim&\lambda_{3}v_{3}^{-1},~~~ \lambda_{4}R_{\tau y\tau
y}g^{\tau\tau}g^{yy}\sim\lambda_{4}v_{4}^{-1}.
\end{eqnarray}
Assume that the $n$-form electric field strength $F_{\tau\rho
yp\cdots q}$ and $m$-form magnetic field strength $H^{(m)}$ satisfy
$F_{\tau\rho yp\cdots q}=e_{1}$ and $H^{(m)}=p_{a}\sqrt{\Omega_{m}}$
at the horizon, where $e_{1}$, $p_{a}$ are constants to be
determined and $\Omega_{m}$ denotes the measure of $m$-dimensional
unit sphere. In the Lagrangian the electric field strength behaves
as
\begin{equation}
\nonumber \sqrt{(g^{\tau\tau}g^{yy}-(g^{\tau
y})^{2})g^{\rho\rho}g^{pp}\cdots g^{qq}}F_{\tau\rho yp\cdots q}\sim
e_{1}v_{1}^{-\frac{3}{2}}.
\end{equation}
Note that no other contributions have any dependence on $v_{1}$.
Then we can rewrite $f_{\lambda}$ as a function of scalars, electric
and magnetic field strengths
\begin{equation}
\label{f1}
f_{\lambda}(u_{s},v_{1},v_{2},e_{1},p_{a})=v_{1}^{\frac{3}{2}}h(u_{s},v_{2},
\lambda_{i}v_{1}^{-1},e_{1}v_{1}^{-\frac{3}{2}},p_{a}),
\end{equation}
where $h$ is a general function and the factor $v_{1}^{\frac{3}{2}}$
comes from $\sqrt{-{\rm det}g}$.

Using~(\ref{f1}), one can easily derive the following equation
\begin{equation}
\label{3eq18} \sum \limits^{4}_{i=1} \left.
\lambda_{i}\frac{\partial f_{\lambda_{i}}}{\partial\lambda_{i}}
\right |_{\lambda_{i}=1}=\frac{3}{2} (f-e_{1}\frac{\partial
f}{\partial e_{1}})-v_{1}\frac{\partial f}{\partial v_{1}}.
\end{equation}
Then the entropy can be reexpressed by substituting~(\ref{3eq18})
into~(\ref{3eq14})
\begin{equation}
\mathcal{S}_{BH}=-2\pi\lambda^{2}[\frac{3}{2}(f-e_{1}\frac{\partial
f}{\partial e_{1}})-\frac{\partial
f}{\partial\lambda_{2}}-\frac{\partial
f}{\partial\lambda_{4}}+\frac{\rho_{+}^{2}\rho_{-}^{2}}{\rho^{2}(\rho_{+}^{2}+\rho_{-}^{2}-\rho^{2})}\frac{\partial
f}{\partial\lambda_{2}}].
\end{equation}
To simplify the above expression, we have to make use of the
following relations, which can be derived from the symmetries of the
Lagrangian,
\begin{equation}
\label{3eq20} \frac{\partial f}{\partial\lambda_{1}}=\frac{\partial
f}{\partial\lambda_{2}},~~~ \frac{\partial
f}{\partial\lambda_{3}}=-\frac{\rho_{+}^{2}\rho_{-}^{2}}{\rho^{2}(\rho_{+}^{2}+\rho_{-}^{2}-\rho^{2})}\frac{\partial
f}{\partial\lambda_{2}},~~~ \frac{\partial
f}{\partial\lambda_{1}}=\frac{\partial
f}{\partial\lambda_{3}}+\frac{\partial f}{\partial\lambda_{4}}.
\end{equation}
With the help of~(\ref{3eq20}), we obtain a simple expression for
the entropy
\begin{eqnarray}
\mathcal{S}_{BH}&=&\pi\lambda^{2}(e_{1}\frac{\partial f}{\partial
e_{1}}-f) \nonumber \\
               &\equiv&\pi\lambda^{2}F.
\end{eqnarray}
Thus we have completed the general derivation for the entropy
function and the entropy can be obtained by extremizing the entropy
function with respect to the moduli
\begin{equation}
\frac{\partial F}{\partial u_{s}}=0,~~~\frac{\partial F}{\partial
v_{i}}=0,~~~i=1,2,
\end{equation}
and then substituting the values of the moduli back into $F$. Note
that the relation $q_{i}=\frac{\partial f}{\partial e_{i}}$ does not
hold any more in the non-extremal case.

Finally, we would like to stress how the entropy function changes if
we rescale the coordinates of the background metric. Note that in
order to obtain the standard $BTZ$ metric, some of the coordinates
have been rescaled in the previous sections, which can be seen
from~(\ref{res1}) and~(\ref{res2}). Suppose we make the following
coordinate rescaling
\begin{equation}
\label{res3} t\rightarrow At
\end{equation}
where $A$ is an arbitrary constant. Since the Lagrangian should be
diffeomorphism invariant, one can see from the definition of
$f$~(\ref{3eq10}) that under the rescaling~(\ref{res3}), one has
\begin{equation}
f\rightarrow Af.
\end{equation}
Thus the entropy function and the entropy behave as
\begin{equation}
\label{3.25}
F\rightarrow AF,~~~\mathcal{S}_{BH}\rightarrow
A\mathcal{S}_{BH}.
\end{equation}
Such transformations will be used in the following sections.
\section{Entropy Function in Ten-dimensional Spacetime}
\label{ten-dim}
We will calculate the entropy function for the two concrete examples
presented in section~\ref{Non-ext} in the present section. We find
that for the five-dimensional non-extremal black hole, the result
agrees with the Bekenstein-Hawking entropy precisely, while for the
four-dimensional non-extremal black hole, the result is different
from the Bekenstein-Hawking entropy by a factor, which can be
understood according to the arguments given at the end of the last
section.

\subsection{Case 1: the 5d non-extremal black hole}
First we determine the near horizon field configuration in ten
dimensional string frame as follows
\begin{eqnarray}
ds^{2}&=&v_{1}(-\frac{(\rho^{2}-\rho_{+}^{2})(\rho^{2}-\rho_{-}^{2})}{\lambda^{2}\rho^{2}}dt^{2}+\frac{\lambda^{2}\rho^{2}}
{(\rho^{2}-\rho_{+}^{2})(\rho^{2}-\rho_{-}^{2})}d\rho^{2} \nonumber
\\
      & &+\rho^{2}(\frac{1}{\lambda}dz-\frac{\rho_{+}\rho_{-}}{\lambda\rho^{2}}dt)^2)+v_{2}(\lambda^{2}d\Omega_{3}^{2}+\frac{r_{1}}{r_{5}}dx_{\|}^{2}),
     \nonumber \\
e^{-2\phi}&=&u_{s},~~~F_{t\rho
z}=e_{1}=\frac{2\rho}{r_{1}^{2}}\frac{v_{1}^{\frac{3}{2}}}{v_{2}^{\frac{7}{2}}},~~~
G_{\theta\varphi\psi}=2r_{5}^{2}\sin^{2}\theta\sin\varphi.
\end{eqnarray}
Using the above field configuration, the effective
action~(\ref{iib}) turns out to be
\begin{equation}
\mathcal{L}=\frac{1}{16\pi
G^{10}_{N}}[u_{s}\frac{6(v_{1}-v_{2})}{r_{1}r_{5}v_{1}v_{2}}+\frac{r_{1}r_{5}}{\rho^{2}v_{1}^{3}}\frac{e_{1}^{2}}{2}
-\frac{2r_{5}}{r_{1}^{3}v_{2}^{3}}]
\end{equation}
and the entropy function becomes
\begin{equation}
F=\frac{VR_{z}r_{1}^{3}\rho}{2g_{s}^{2}r_{5}}[u_{s}\frac{6(v_{2}-v_{1})}{r_{1}r_{5}v_{1}v_{2}}+\frac{r_{1}r_{5}}{\rho^{2}v_{1}^{3}}\frac{e_{1}^{2}}{2}
+\frac{2r_{5}}{r_{1}^{3}v_{2}^{3}}],
\end{equation}
where we have set $l_{s}=1$ so that $G^{10}_{N}=8\pi^{6}g_{s}^{2}$.
Substituting the value of $e_{1}$ and solving the equations
\begin{equation}
\frac{\partial F}{\partial u_{s}}=0,~~~\frac{\partial F}{\partial
v_{i}}=0,~~~i=1,2,
\end{equation}
we have
\begin{equation}
\label{result}
u_{s}=\frac{r_{5}^{2}}{r_{1}^{2}},~~~v_{1}=1,~~~v_{2}=1,
\end{equation}
which gives the correct values of the moduli fields. Put the
solution~(\ref{result}) back into $F$, we obtain
\begin{equation}
F=\frac{2VR_{z}}{g_{s}^{2}}\rho.
\end{equation}
Furthermore,
\begin{eqnarray}
\mathcal{S}_{BH}&=&\pi\lambda^{2}F\mid_{\rho=\rho_{+}} \nonumber \\
                &=&\frac{2\pi
                VR_{z}}{g_{s}^{2}}r_{1}r_{5}r_{H}\cosh\alpha_{m} \nonumber
                \\
                &=&S_{BH},
\end{eqnarray}
which is just the black hole entropy in the decoupling limit. Note
that here we have used the fact that in this limit,
$\sinh\alpha_{1,5}\approx\cosh\alpha_{1,5}$.
\subsection{Case 2: the 4d non-extremal black hole}
The entropy function for four-dimensional black hole can be
calculated in a similar way. First we write down the near horizon
field configuration in ten dimensional string frame
\begin{eqnarray}
\label{4.8}
ds^{2}&=&v_{1}[-\frac{1}{(r_{2}r_{6})^{\frac{1}{2}}}\frac{(\rho^{2}-\rho_{+}^{2})(\rho^{2}-\rho_{-}^{2})}{\rho^{2}}dt^{2}
+\frac{4r_{5}(r_{2}r_{6})^{\frac{1}{2}}\rho^{2}}{(\rho^{2}-\rho_{+}^{2})(\rho^{2}-\rho_{-}^{2})}d\rho^{2}
 \nonumber \\
      & &+\frac{\rho^{2}}{(r_{2}r_{6})^{\frac{1}{2}}}(dz-\frac{\rho_{+}\rho_{-}}{\rho^{2}}dt)^2]
+v_{2}[r_{5}(r_{2}r_{6})^{\frac{1}{2}}d\Omega_{2}^{2} \nonumber \\
      & &\frac{r_{5}}{(r_{2}r_{6})^{\frac{1}{2}}}dx_{2}^{2}+(\frac{r_{2}}{r_{6}})^{\frac{1}{2}}dx_{\|}^{2}],
    \nonumber \\
e^{-2\phi}&=&u_{s},~~~F_{tz\rho2}=e_{1}=\frac{\rho}{r_{2}}\frac{v_{1}^{\frac{3}{2}}}{v_{2}^{\frac{5}{2}}},
 \nonumber \\
H_{2\theta\varphi}&=&-\frac{1}{2}r_{5}\sin\theta,~~~G_{\theta\varphi}=-\frac{1}{2}r_{6}\sin\theta.
\end{eqnarray}
Under the above field configuration, the effective
action~(\ref{iia}) becomes
\begin{equation}
\mathcal{L}=\frac{1}{16\pi
G^{10}_{N}}[u_{s}(\frac{4v_{1}-3v_{2}}{2r_{5}(r_{2}r_{6})^{\frac{1}{2}}v_{1}v_{2}}-\frac{1}{2r_{5}(r_{2}r_{6})^{\frac{1}{2}}v_{2}^{3}})
-\frac{r_{6}}{2r_{2}r_{5}^{2}v_{2}^{3}}+\frac{e_{1}^{2}r_{2}r_{6}}{2\rho^{2}r_{5}^{2}v_{1}^{3}v_{2}}]
\end{equation}
and the entropy function turns out to be
\begin{equation}
F=\frac{4\rho
VR_{2}R_{z}}{g_{s}^{2}}\frac{r_{2}v_{2}^{\frac{1}{2}}}{2(r_{2}r_{6})^{\frac{3}{2}}v_{1}^{\frac{3}{2}}}
\{\frac{(r_{2}r_{6})^{\frac{3}{2}}v_{1}^{3}}{r_{2}v_{2}^{3}}+v_{1}^{2}[r_{2}^{\frac{1}{2}}r_{6}^{\frac{3}{2}}v_{1}v_{2}
+r_{2}r_{5}u_{s}(v_{1}-4v_{1}v_{2}^{2}+3v_{2}^{3})]\},
\end{equation}
Solving the equations
\begin{equation}
\frac{\partial F}{\partial u_{s}}=0,~~~\frac{\partial F}{\partial
v_{i}}=0,~~~i=1,2,
\end{equation}
we arrive at the correct attractor values of the moduli fields
\begin{equation}
\label{res}
u_{s}=\frac{r_{6}^{\frac{3}{2}}}{r_{2}^{\frac{1}{2}}r_{5}},~~~
v_{1}=1,~~~v_{2}=1.
\end{equation}
Substituting the solutions~(\ref{res}) back into $F$, we obtain
\begin{equation}
F=\frac{4\rho VR_{2}R_{z}}{g_{s}^{2}}.
\end{equation}
Finally,
\begin{eqnarray}
\label{4dentropy}
\mathcal{S}_{BH}&=&\pi\lambda^{2}F\mid_{\rho=\rho_{+}}
           \nonumber \\
                &=&\frac{16\pi
                VR_{2}R_{z}}{g_{s}^{2}}(r_{2}r_{6})^{\frac{1}{2}}r_{5}\rho_{+}
                \nonumber
                \\
                &=&2\sqrt{r_{5}}S_{BH},
\end{eqnarray}
where we have use the fact that
$\sinh\alpha_{2,5,6}\approx\cosh\alpha_{2,5,6}$ in the near horizon
region and the result does not agree with the Bekenstein-Hawking
entropy by a  factor $2 \sqrt{r_5}$. Note that one has to make a
rescaling transformation (\ref{res2}) in order to transform the part
spanned by coordinates $(t, \rho, z)$ in (\ref{4.8}) to be a
standard BTZ metric. Further note that one has the transformation
relation (\ref{3.25}) due to the rescaling (\ref{res3}). Thus the
result (\ref{4dentropy}) indeed gives us the entropy of 4d black
holes in ten dimensional string frame.

\section{Entropy Function in Lower Dimensions}
\label{lower}
It is well known that after dimensional reduction, the extremal
black hole in Type II string theory has $AdS_{2}$ as part of its
near horizon geometry rather than $AdS_{3}$. It has already been
noticed in~\cite{ag} and~\cite{cp} that although the entropy
function could give the correct entropy in lower dimensions, not all
the moduli fields could take definite values. In this section we
first do the dimensional reduction down to six and five dimensions,
keeping the BTZ part of the near horizon metric invariant, then we
 find that the same results for the entropy can be obtained
while some of the moduli fields do not take definite values.
\subsection{Case 1: the 5d non-extremal black hole}
We do the dimensional reduction on $x_{\|}$ and obtain a
six-dimensional black string with near horizon geometry $BTZ\times
S^{3}$. The near horizon field configuration
\begin{eqnarray}
ds^{2}&=&v_{1}(-\frac{(\rho^{2}-\rho_{+}^{2})(\rho^{2}-\rho_{-}^{2})}{\lambda^{2}\rho^{2}}dt^{2}+\frac{\lambda^{2}\rho^{2}}
{(\rho^{2}-\rho_{+}^{2})(\rho^{2}-\rho_{-}^{2})}d\rho^{2}\nonumber
\\
      & &+\rho^{2}(\frac{1}{\lambda}dz-\frac{\rho_{+}\rho_{-}}{\lambda\rho^{2}}dt)^2)+v_{2}\lambda^{2}d\Omega_{3}^{2},\\\nonumber
e^{-2\phi}&=&u_{s},~~~e^{2\psi}=u_{T} \nonumber \\
F_{t\rho
z}&=&e_{1}=\frac{2\rho}{u_{T}r_{5}^{2}}\frac{v_{1}^{\frac{3}{2}}}{v_{2}^{\frac{3}{2}}},~~~
G_{\theta\varphi\psi}=2r_{5}^{2}\sin^{2}\theta\sin\varphi,
\end{eqnarray}
where $e^{2\psi}$ stands for the single moduli for $T^{4}$.

The six-dimensional effective Lagrangian, which can be obtained by
the standard procedure ( see e.g.~\cite{klp}), becomes
\begin{equation}
\mathcal{L}=\frac{1}{16\pi G^{6}_{N}}\sqrt{-{\rm
det}g^{(6)}}e^{2\psi}
\{e^{-2\phi}[R^{(6)}+4(\nabla\phi)^2]-\frac{1}{2}\sum\limits_{n}\frac{1}{n!}F^{2}_{n}\},
\end{equation}
where the superscript stands for that the quantities stay in six
dimensions. Using the near horizon field configuration and the
effective action, the entropy function is expressed as
\begin{equation}
F=\frac{VR_{z}\rho
r_{1}r_{5}}{2g_{s}^{2}}v_{1}^{\frac{3}{2}}v_{2}^{\frac{3}{2}}u_{T}
[u_{s}\frac{6(v_{2}-v_{1})}{r_{1}r_{5}v_{1}v_{2}}+\frac{r_{1}r_{5}}{\rho^{2}v_{1}^{3}}\frac{e_{1}^{2}}{2}
+\frac{2r_{5}}{r_{1}^{3}v_{2}^{3}}].
\end{equation}

Solving the equations after substituting the value of $e_{1}$ into
$F$,
\begin{equation}
\frac{\partial F}{\partial u_{s}}=0,~~~\frac{\partial F}{\partial
u_{T}}=0,~~~\frac{\partial F}{\partial v_{i}}=0,~~~i=1,2,
\end{equation}
we obtain
\begin{equation}
v_{1}=v_{2}=v,~~~u_{s}=\frac{r_{5}^{2}}{r_{1}^{2}v^{2}},~~~u_{T}=\frac{r_{1}^{2}}{r_{5}^{2}},
\end{equation}
where $v$ is an arbitrary constant. Finally, after substituting the
solutions back into $F$, we get
\begin{eqnarray}
\mathcal{S}_{BH}&\equiv&\pi\lambda^{2}F\mid_{\rho=\rho_{+}}
     \nonumber \\
                &=&\frac{2\pi
                VR_{z}r_{1}r_{5}\rho_{+}}{g_{s}} \nonumber \\
                &=&S_{BH},
\end{eqnarray}
which is again the entropy of 5d non-extremal black holes.
\subsection{Case 2: the 4d non-extremal black hole}
Similarly, we do the dimensional reduction on $x^{2}\times x_{\|}$
and obtain a five-dimensional black string with near horizon
geometry $BTZ\times S^{2}$. The near horizon field configuration is
\begin{eqnarray}
ds^{2}&=&v_{1}[-\frac{1}{(r_{2}r_{6})^{\frac{1}{2}}}\frac{(\rho^{2}-\rho_{+}^{2})(\rho^{2}-\rho_{-}^{2})}{\rho^{2}}
dt^{2}+\frac{4r_{5}(r_{2}r_{6})^{\frac{1}{2}}\rho^{2}}{(\rho^{2}-\rho_{+}^{2})(\rho^{2}-\rho_{-}^{2})}d\rho^{2}
 \nonumber \\
      & &+\frac{\rho^{2}}{(r_{2}r_{6})^{\frac{1}{2}}}(dz-\frac{\rho_{+}\rho_{-}}{\rho^{2}}dt)^2]
+v_{2}r_{5}(r_{2}r_{6})^{\frac{1}{2}}d\Omega_{2}^{2}, \nonumber \\
e^{-2\phi}&=&u_{s},~~~e^{2\psi}=u_{T},~~~e^{\frac{\psi_{1}}{2}}=u_{1},
\nonumber
\\ F^{(5)}_{tz\rho}&=&e_{1}=\frac{\rho
u_{1}}{u_{T}}\frac{r_{2}^{\frac{1}{4}}}{r_{5}^{\frac{1}{2}}r_{6}^{\frac{3}{4}}}\frac{v_{1}^{\frac{3}{2}}}{v_{2}},~~~
H^{(5)}_{\theta\varphi}=-\frac{1}{2}r_{5}\sin\theta,~~~G_{\theta\varphi}=-\frac{1}{2}r_{6}\sin\theta,
\end{eqnarray}
where $e^{2\psi}$ and $e^{\frac{\psi_{1}}{2}}$ denote the single
moduli for $T^{4}$ and $S^{1}$ respectively.

The effective Lagrangian in five dimensions can be expressed as
\begin{equation}
\mathcal{L}=\frac{1}{16\pi G^{5}_{N}}\sqrt{-{\rm
det}g^{(5)}}e^{2\psi}e^{\frac{\psi_{1}}{2}}[e^{-2\phi}(R^{(5)}-e^{-\psi_{1}}H^{(5)2})
-G^{2}-\frac{1}{3}e^{-\psi_{1}}F^{(5)2}],
\end{equation}
where the superscript signifies that the quantities stay in five
dimensions and the 2-form magnetic field strength $H^{(5)}$ as well
as the 3-form electric field strength $F^{(5)}$ originate from the
ten-dimensional field strengths $H_{2\theta\varphi}$ and $F_{tz\rho
2}$. Note that we have omitted the terms involving the covariant
derivatives of the scalar fields because they are set to be
constants at the horizon.

We can work out the five-dimensional entropy function by making use
of the above effective action and near horizon field configuration
\begin{eqnarray}
F&=&\frac{4VR_{2}R_{z}\rho
r_{5}^{\frac{3}{2}}(r_{2}r_{6})^{\frac{1}{4}}v_{1}^{\frac{3}{2}}v_{2}u_{T}u_{1}}{g_{s}^{2}}
 \left [u_{s}(\frac{3v_{2}-4v_{1}}{2r_{5}(r_{2}r_{6})^{\frac{1}{2}}v_{1}v_{2}}+\frac{1}{2u_{1}^{2}r_{2}r_{6}v_{2}^{2}})
  \right. \nonumber
\\
 & & \left. +\frac{r_{6}}{2r_{2}r_{5}^{2}v_{2}^{2}}+\frac{e_{1}^{2}(r_{2}r_{6})^{\frac{1}{2}}}{2u_{1}^{2}
  \rho^{2}r_{5}v_{1}^{3}} \right ].
\end{eqnarray}
After substituting the value of $e_{1}$ we can solve the equations
\begin{equation}
\frac{\partial F}{\partial u_{i}}=0,~~i=s,T,1,~~~~\frac{\partial
F}{\partial v_{j}}=0,~~j=1,2,
\end{equation}
and obtain
\begin{eqnarray}
v_{1}&=&v,~~~~v_{2}=v,\nonumber \\ u_{s}&=
&\frac{r_{6}^{\frac{3}{2}}}{r_{5}r_{2}^{\frac{1}{2}}v},
~~~u_{T}=\frac{r_{2}}{r_{6}},~~~u_{1}=\frac{r_{5}^{\frac{1}{2}}}{(r_{2}r_{6})^{\frac{1}{4}}v^{\frac{1}{2}}},
\end{eqnarray}
where $v$ is an arbitrary constant, once again.

The entropy can be obtained after substituting the solution back
into $F$
\begin{eqnarray}
\mathcal{S}_{BH}&\equiv&\pi\lambda^{2}F\mid_{\rho=\rho_{+}}
  \nonumber \\
                &=&\frac{16\pi
                VR_{2}R_{z}(r_{2}r_{6})^{\frac{1}{2}}r_{5}\rho_{+}}{g_{s}^{2}} \nonumber
                \\
                &=&2\sqrt{r_{5}}S_{BH}.
\end{eqnarray}
Here the factor $2\sqrt{r_5}$ appears again. The reason is the same
as the one discussed in the previous section.
\section{Summary and Discussion}
\label{Summary}
The entropy function formalism proposed by Sen is an efficient way
to calculate the entropy of a black hole with $AdS_{2}$ as part of
the spacetime geometry. However, as far as we know, most of the work
have been dealing with extremal black holes. In this paper we show
that for some non-extremal black holes in string theory with $BTZ$
as part of the near horizon geometry, the entropy function formalism
also works very well and can reproduce the Bekenstein-Hawking
entropy both in ten dimensions and lower dimensions.  Thus our work
generalizes the entropy function formalism to certain non-extremal
black holes and we expect that it might also work for other
non-extremal black objects, such as black $p$-branes.

We notice that a relevant issue was presented recently
in~\cite{dst}, which describes how to apply the entropy function
formalism to near-extremal case. They argued that in order to deal
with the runaway behavior of the entropy function, one has to
introduce a slight amount of non-extremality on the black hole side.
The non-extremality parameter $\epsilon$ truncates the infinite
throat of $AdS_{2}$ in to a finite size, thus the near horizon
geometry is no longer $AdS_{2}\times S^{d-2}$. But for sufficiently
large charges and small $\epsilon$ there will be a region in the
black hole spacetime where the geometry is approximately
$AdS_{2}\times S^{d-2}$, and one can use the entropy function
formalism to calculate the entropy in this region. However, in our
examples the near horizon geometry $BTZ$ do not rely on the
near-extremal limit, but the entropy function formalism still works.
It would be interesting to study the relations between the two
approaches extensively.

Recently, an intuitional explanation of the black hole
attractor/non-attractor behavior has been proposed in~\cite{kal},
which states that the attractor/non-attractor behavior is closely
related to the near horizon geometry. For extremal black holes with
$AdS_{2}$ near horizon geometry, the physical distance from a finite
radius coordinate $r_{0}$ to the horizon turns out to be infinite,
while for non-extremal black holes the distance remains finite. It
is clear that the infinite physical distance is crucial to allow a
scalar field to forget its initial conditions while in non-extremal
case the field only has finite ``time'' until it reaches the
horizon. In our examples, the physical distance from a finite radial
coordinate $\rho_{0}$ to the outer horizon becomes
\begin{eqnarray}
d&=&\int^{\rho_{0}}_{\rho_{+}}\sqrt{g_{\rho\rho}}d\rho\\\nonumber
 &=&\lambda\log(\sqrt{\rho^{2}-\rho_{+}^{2}}+\sqrt{\rho^{2}-\rho_{-}^{2}})\mid^{\rho=\rho_{0}}_{\rho=\rho_{+}}\\\nonumber
 &=&\lambda[\log(\sqrt{\rho_{0}^{2}-\rho_{+}^{2}}+\sqrt{\rho_{0}^{2}-\rho_{-}^{2}})-\log(\sqrt{\rho_{+}^{2}-\rho_{-}^{2}})],
\end{eqnarray}
which turns out to be finite. However, in certain cases, the scalar
fields considered here do exhibit some ``attractor'' behavior, that
is, the values at the horizon can be determined by extremizing the
entropy function. So it is worth investigating this phenomenon
thoroughly.

\acknowledgments DWP would like to thank Hua Bai, Li-Ming Cao and
Jian-Huang She for useful discussions and kind help. The work was
supported in part by a grant from Chinese Academy of Sciences, by
NSFC under grants No. 10325525 and No. 90403029.
\end{document}